\documentclass[twocolumn]{autart}    

\usepackage{graphicx}    
\usepackage{epsfig} 
\usepackage{mathptmx} 
\usepackage{times} 
\usepackage{amsmath} 
\usepackage{amssymb}  
\usepackage{graphicx} 
\usepackage{makecell} 
\usepackage{latexsym,amsfonts,amsbsy}
\usepackage{comment}
\usepackage{algorithm}
\usepackage[noend]{algpseudocode}

\usepackage{xcolor}

\usepackage{flushend}    
\usepackage{comment}
\usepackage{here}
\usepackage{mathtools}
\usepackage{enumerate}
\usepackage{nccmath}
\usepackage{booktabs}
\usepackage{multirow}
\usepackage{url}

\newtheorem{theorem}{Theorem}

\newtheorem{remark}{Remark}

\algrenewcommand\algorithmicrequire{\textbf{Input:}}
\algrenewcommand\algorithmicensure{\textbf{Output:}}

\begin{document}

\begin{frontmatter}

\title{Markov Clustering based Fully Automated Nonblocking Hierarchical Supervisory Control of Large-Scale Discrete-Event Systems}


\author[Yingying]{Yingying Liu}\ead{yingyingliu611@163.com},    
\author[Kai]{Zhaojian Cai}\ead{zhaojian.cai@c.info.eng.osaka-cu.ac.jp},               
\author[Kai]{Kai Cai}\ead{cai@omu.ac.jp}  

\address[Yingying]{School of Information Engineering,
         Northwest A$\&$F University, 712100 Xianyang, China}  
\address[Kai]{Department of Core Informatics, Osaka Metropolitan University, 5588585 Osaka, Japan}             


\begin{abstract}                          
In this paper we revisit the {\it abstraction-based approach} to synthesize a hierarchy of decentralized supervisors and coordinators for nonblocking control of large-scale discrete-event systems (DES), and augment it with a new clustering method for automatic and flexible grouping of relevant components during the hierarchical synthesis process. This method is known as {\it Markov clustering}, which not only automatically performs grouping but also allows flexible tuning the sizes of the resulting clusters using a single parameter.  Compared to the existing abstraction-based approach that lacks effective grouping method for general cases, our proposed approach based on Markov clustering provides a fully automated and effective hierarchical synthesis procedure applicable to general large-scale DES. Moreover, it is proved that the resulting hierarchy of supervisors and coordinators collectively achieves global nonblocking (and maximally permissive) controlled behavior under the same conditions as those in the existing abstraction-based approach. Finally, a benchmark case study is conducted to empirically demonstrate the effectiveness of our approach.
\end{abstract}

\end{frontmatter}

\section{Introduction}
 The design of supervisors for large-scale discrete-event systems (DES)  has become a challenge as they comprise growing numbers of components to control. Some representative methods have been proposed, such as centralized control\cite{Komenda2023}, heterarchical (combination of decentralized \cite{Deng2021,9361277} and hierarchical\cite{Pasquale2020}) control, and  distributed control\cite{Zhang2020} for large-scale DES.
In this paper, we revisit the {\em abstraction-based approach} to synthesize a hierarchy of decentralized supervisors and coordinators for large-scale DES \cite{Wong1998,Feng2008,Schmidt2008,Schmidt2011,su2012maximally,Mohajerani2015}. 
In a nutshell, this approach (i) computes decentralized (or modular) supervisors to enforce individual specifications, (ii) abstracts these supervisors by projections, and (iii) ensures nonblocking joint behavior of the abstracted supervisors by designing coordinators (a coordinator is a higher-level supervisor whose sole role is to remove blocking states). The strength of this approach lies in that under certain conditions on the projections  (language-based  \cite{Goorden2021,Wong1998,Feng2008,Schmidt2008,Schmidt2011} or state-based \cite{Ju2021,su2012maximally,Mohajerani2015}) used in step~(ii), the coordinators designed for abstractions in step~(iii) ensures nonblocking behavior of the overall system (further conditions on the projections can moreover ensure maximal permissiveness).

 However, for large DES with many components and specifications, it is not uncommon that there is a large number of decentralized supervisors in step~(i), and thus the same large number of abstracted supervisors in step~(ii). This may well render the designing of coordinators in step~(iii) computationally infeasible in one-shot.
In such a case, steps~(ii) and (iii) are instead carried out in a {\it hierarchical} (i.e. multi-level) manner: Group decentralized supervisors into clusters of manageable sizes, design coordinators (if needed) for individual clusters, and abstract the clusters by projections; in further higher levels, treat the abstracted clusters as supervisors, and repeat the above process of grouping and coordinator design until a single (top-level) coordinator is resulted (or top-level nonblockingness is verified). This hierarchical abstraction-based approach also ensures nonblocking (and maximally permissive) behavior of the overall system under the same conditions on the projections  \cite{Feng2008,su2012maximally}. Note that although not explicit, different methods of creating abstractions presented in \cite{Wong1998,Schmidt2008,Schmidt2011,Mohajerani2015} can be readily adapted to the above mentioned hierarchical approach.

Although this hierarchical (divide-and-conquer type) approach is effective in addressing computational feasibility, the approach brings out a new issue: {\em How to group the decentralized supervisors, or the abstracted clusters on higher levels?}
In fact, better groupings typically result in more efficient coordinator synthesis. In \cite{Feng2008,FengWonCai2009}, {\em ad hoc} good groupings are found for specific case studies by exploiting the corresponding systems'  structural features. Using such handcrafted groupings is, however, case-dependent and hence prevents the hierarchical abstraction-based approach from being fully automatic when applied to arbitrary DES. 

What is needed, then, is an {\it algorithmic grouping recipe} applicable for general cases. Two such recipes exist in the literature. The first one is based on {\em shared event} \cite{Feng2008}: namely, two supervisors/abstractions are grouped together as long as they share a common event. Although straightforward, this way of grouping often results in large numbers of supervisors/abstractions grouped into the same cluster, which counterproductively renders the coordinator design for the cluster computationally inefficient.  
The second recipe is {\it sequential grouping} \cite{su2012maximally}. This recipe {\em prescribes} a priority order on the supervisors (indeed any components to be grouped), and performs along that fixed order both grouping and abstraction. The effectiveness of this sequential recipe depends highly on the prescribed order, and how to choose a good order is an open problem. 

In view of above-mentioned state-of-the-art of the abstraction-based approach, in this paper we propose a new algorithmic grouping method which is more effective and general than those based on shared events or sequential orders. 
Our grouping method employs the algorithm known as {\em Markov clustering} \cite{MCL}, which not only automatically performs grouping but also allows flexible tuning the sizes of the resulting clusters using a single parameter. The latter feature is convenient if bigger or smaller clusters are sought. To integrate Markov clustering into the hierarchical abstraction-based approach, we adopt the {\em dependency structure matrix} (DSM) \cite{goorden2019structuring} to encode the relationship between plant components and the relevant entities to be clustered. Specifically, DSM is first used to encode the relation between plant components and decentralized supervisors (or specifications enforced by these supervisors), so as to group the supervisors into clusters. Then in higher levels, DSM is used to encode the relation between plant components and abstracted clusters for grouping these abstractions into clusters.

The contributions of our  Markov clustering based hierarchical abstraction approach are summarized below. 
\begin{itemize}
    \item First, compared to the existing abstraction-based approach (including the interface-based method \cite{Leduc2005} and the supervisor localization method \cite{cai2016}) with handcrafted, shared-event based, or sequential-order based clustering, our new approach with Markov clustering provides a fully automated, effective, and flexible hierarchical synthesis procedure for large-scale DES, with no need of knowing or analyzing system structures.
    \item Second, it is proved that the resulting hierarchy of supervisors and coordinators by our approach collectively achieves global nonblocking (and maximally permissive) controlled behavior under the same conditions on the projections as in \cite{Feng2008,Schmidt2011} (i.e. the correct-by-construction property of the abstraction-based approach is preserved).  This is advantageous as compared to \cite{goorden2019structuring}.
    \item Third, our approach additionally provides flexibility in tuning the cluster sizes by adjusting a single parameter used in Markov clustering. This may be useful in adapting our approach to specific computational capabilities by choosing smaller or bigger clusters.
    \item  Fourth, our approach is modular which can result in smaller and thus more comprehensible decentralized supervisors/coordinators than centralized state-tree-structure based approach \cite{Ma2006}. 
\end{itemize}
Overall, these benefits of our proposed approach can potentially make the abstraction-based approach a practical technology for engineers and industries: 
 With execution of one command,
a set of correct supervisors and coordinators can be computed for any systems and specifications; also with adjusting one parameter, different-size clusters may be obtained to be adapted for different needs.

Our use of Markov clustering is inspired by the work in \cite{goorden2019structuring}, which develops a DSM-based approach to transform a given set of plant components and specifications modeled as extended finite automata into a tree-structured multi-level DES, and then the multi-level supervisory control synthesis \cite{komenda2013multilevel} is applied.
There are three main differences between \cite{goorden2019structuring} and our work. First, the targets to be clustered are different. In \cite{goorden2019structuring}, plant components are to be clustered and rendered into a tree structure. In this paper, we cluster decentralized supervisors as well as higher-level abstractions. This difference is reflected in using DSM to encode different relationships. Second, while clustering in \cite{goorden2019structuring} is used once for plant components, we use clustering multiple times in a hierarchical fashion -- this is to be compatible with the abstraction-based approach.  Last and most importantly, the supervisors synthesized in \cite{goorden2019structuring} based on the multi-level approach \cite{komenda2013multilevel} do not generally ensure global nonblocking behavior; by contrast, nonblockingness is ensured by our approach based on abstractions with certain properties imposed.

We note that from a different perspective, an approach to address global nonblockingness (and maximal permissiveness) is developed in \cite{Goorden2021RCNMS}. There the authors first introduce ``controllable and nonblocking modular supervisors properties (CNMSP)'', and show that the global nonblocking verification or coordinator synthesis is not needed if the supervisory control problem satisfies the CNMSP. In case the CNMSP fails to hold, namely global nonblocking verification and coordinator synthesis may be needed, our proposed approach can be used.

\section{PRELIMINARIES AND PROBLEM STATEMENT}\label{sec:preliminaries}
The DES plant to be controlled is modeled by a generator
$\textbf{G}=(Q,\Sigma,\delta,q_{0},Q_{m})$,
where $Q$ is the finite state set, $\Sigma=\Sigma_{c}\dot{\cup} \Sigma_{u}$ is the finite event set which is partitioned into two subsets --
the controllable event subset $\Sigma_{c}$ and the uncontrollable event subset $\Sigma_{u}$.
The function $\delta: Q\times \Sigma\rightarrow Q$ is the (partial) state transition function,
$q_{0}\in Q$ is the initial state, and $Q_{m}\subseteq Q$ is the set of marker states.
In the usual way, we extend $\delta$ such that $\delta: Q\times \Sigma^{*}\rightarrow Q$,
and write $\delta(q,s)!$ to mean that $\delta(q,s)$ is defined, where $q\in Q$ and $s\in \Sigma^{*}$. 
A string $s_1\in \Sigma^{*}$ is a {\it prefix} of another string $s\in \Sigma^{*}$, written $s_1\leq s$,
if there exists $s_2\in \Sigma^{*}$ such that $s_1s_2$ = $s$.
The {\it prefix closure} of a language $L \subseteq \Sigma^*$, written $\overline{L}$,
is $\overline{L}: = \{s_1\in\Sigma^{*} \mid (\exists s\in L) s_1\leq s\}$.
A language $L$ is \emph{closed} if $L = \overline{L}$.


The closed behavior of \textbf{G} is the set of all strings that can be generated by \textbf{G}: $L(\textbf{G}):=\{s \in \Sigma^*|\delta(q_0,s)!\}.$
As defined $L(\textbf{G})$ is closed. On the other hand, the marked behavior of \textbf{G} is the subset of strings that can reach a marker state:
$L_m(\textbf{G}):=\{s \in L(\textbf{G})|\delta(q_0,s) \in Q_m \}\subseteq L(\textbf{G}).$
If $L(\textbf{G}) = \overline{L_m(\textbf{G})}$, we say that {\bf G} is nonblocking.  A language $L$ is {\rm controllable}
with respect to \textbf{G} if $\bar{L}\Sigma_u\cap L(\textbf{G})\subseteq \bar{L}$,  where $\bar{L}\Sigma_u=\{s\sigma\in \Sigma^*| s\in \bar{L}, \sigma\in \Sigma_u\}$. We denote by ${\it C}(L)$ the family of all controllable sublanguages of $L$, which contains a (unique) supremal element $\textrm{sup}\textit{C}(L):=\cup \{L'|L'\in \textit{C}(L)\}$ \cite{wonham2019supervisory}. 

\noindent {\bf Nonblocking Hierarhical Control Problem}

Let the plant to be controlled consist of $N$ $(> 1)$ component $\textbf{G}_k =(Q_k,\Sigma_k,\delta_k, q_{0,k},Q_{m,k}), k = 1,\ldots,N$.
Each component's event set $\Sigma_k$ is partitioned into a controllable subset $\Sigma_{c,k}$ and an
uncontrollable subset $\Sigma_{u,k}$, i.e., $\Sigma_k=\Sigma_{c,k}\dot{\cup} \Sigma_{u,k}$.
Also consider (local) specification languages $\textit{E}_i\subseteq\Sigma^{\ast}_{\textit{E}_i}(i\in\it{I})$,
where $\Sigma_{\textit{E}_i}$ 
is the event set over which $\textit{E}_i$ is defined and $\it{I}$ is an index set. 
 For large-scale DES, both $N$ and $|I|$ are large.

For each specification $\textit{E}_i$, the corresponding (local) plant $\textbf{G}_{\textit{E}_i}$ is the {\em synchronous product} \cite{wonham2019supervisory} of the components that share events with $\Sigma_{\textit{E}_i}$:
\begin{eqnarray}\label{gei}
\textbf{G}_{\textit{E}_i} := ||\{\textbf{G}_k | \Sigma_k \cap \Sigma_{{\it E_i}} \neq\emptyset\}.
\end{eqnarray}
Here $\|$ denotes synchronous product.
Then we synthesize a nonblocking (and maximally permissive) {\it decentralized} {\it supervisor} $\textbf{SUP}_i$ to enforce $E_i$ on $\textbf{G}_{\textit{E}_i}$, namely
\begin{eqnarray}\label{dec.sup}
L_m({\bf SUP}_i)=\textrm{sup}\textit{C}(L_m(\textbf{G}_{E_i})|| E_i).
\end{eqnarray}
In general, the joint behavior of the decentralized supervisors fails to be nonblocking, 
i.e. 
\begin{align} \label{eq:conflict}
||_{i\in I} L(\textbf{SUP}_i)^{\supset}_{\neq} \overline{||{}_{i\in I}L_m(\textbf{SUP}_i)}.
\end{align}
In this case we say that the decentralized supervisors are conflicting. To resolve the conflicts, 
additional higher-level supervisors $\textbf{CO}_j$ ($j \in J$, $J$ some index set) need to be designed. These supervisors are called {\em coordinators}.

Therefore the Nonblocking Hierarchical Control Problem is formulated in the following.

{\bf Problem~1:}
Consider the plant to be controlled consist of $N$ $(> 1)$ component $\textbf{G}_k =(Q_k,\Sigma_k,\delta_k, q_{0,k},Q_{m,k})$ and (local) specification languages $\textit{E}_i\subseteq\Sigma^{\ast}_{\textit{E}_i}$. 
To find the coordinators $\textbf{CO}_j$ ($j \in J$) such that  the decentralized supervisors  $\textbf{SUP}_i$ to enforce $E_i$ on $\textbf{G}_{\textit{E}_i}$ are non-conflicting, i.e.,
\begin{eqnarray}\label{co}
& \overline{\left(||_{i\in I} L_m(\textbf{SUP}_i) \right) \ || \ \left( ||_{j\in J} L_m(\textbf{CO}_j) \right)}
\notag \\
& =\left(||_{i\in I} L(\textbf{SUP}_i) \right) \ || \ \left( ||_{j\in J} L(\textbf{CO}_j) \right).
\end{eqnarray}

\section{Abstraction-based Approach with Markov Clustering}
\label{chap:method}

In this section, we present our proposed abstraction-based approach with Markov clustering, and show that this approach is a solution to the Nonblocking Hierarchical Control Problem formulated at the end of the previous section. 

 Our proposed approach is presented as {\bf Algorithm~1}. Recall that we are given plant components $\textbf{G}_k$ ($k \in \{1, \dots,N\})$ and specifications $E_i$ ($ i \in I$); these are the inputs of the algorithm. The inputs also include two parameters $\alpha, \beta$ for Markov clustering, which will be explained below. The outputs of the algorithm are a set of decentralized supervisors and another set of coordinators.  We will provide a detailed explanation of each step in {\bf Algorithm~1} in the following.

\begin{algorithm}
\caption{Markov Clustering based Nonblocking Hierarchical Supervisory Synthesis Algorithm}\label{alg.2}
\begin{algorithmic}[1]
\Require Plant components 
$\textbf{G}_k$ ($k \in \{1, \dots,N\})$, specifications $E_i$ ($ i \in I$), and Markov clustering parameters $\alpha, \beta$.
\Ensure $\{ \textbf{SUP}_i \mid i \in I\}$ and $\{ {\bf CO}_j \mid j \in J\}$.
\For{$i \in I$} 
\State $\textbf{G}_{\textit{E}_i} = ||\{\textbf{G}_k | \Sigma_k \cap \Sigma_{{\it E_i}} \neq\emptyset\}$
\State $L_m({\bf SUP}_i)=\textrm{sup}\textit{C}(L_m(\textbf{G}_{E_i}|| E_i))$
\EndFor
\State $\mathcal{T} := \{{\bf SUP}_i \mid i \in I\}$
\If{$|\mathcal{T}| \leq 2$}
\State $\mathcal{C} = \{ C_1 \}$ where $C_1 := \mathcal{T}$
\Else 
\State $\mathcal{C}=$MarkovClustering$(\mathcal{T}, \alpha, \beta)$
\EndIf
\While{$|\mathcal{C}|\geq 2$}
\For{$l  = 1$ to $|\mathcal{C}|$}
\If{$||_{T_i \in C_l} 
L(T_i) =  \overline{||{}_{T_i \in C_l}L_m(T_i)}$, where $T_i \in \mathcal{T}$ and $C_l \in \mathcal{C}$}
\State $ {\bf CLU}_l = 
||_{T_i \in C_l} T_i$
\Else
\State $\Sigma_l:=\cup_{T_i \in C_l}\Sigma_{T_i}$, where $\Sigma_{T_i}$ is $T_i$'s event set
\State $L_m({\bf CO}_l)=\sup C( (||{}_{T_i \in C_l}L_m(T_i)) || \Sigma^*_l )$
\State $L_m({\bf CO}_l)=\sup C( (||{}_{T_i \in C_l}L_m(T_i)) || \Sigma^*_l )$
\EndIf
\State $L_m(\textbf{ABS}_l)=\textit{P}_l \left( L_m(\textbf{CLU}_l)\right)$,
where $\textit{P}_l : \Sigma^{\ast}_{\textbf{CLU}_l} \rightarrow \Sigma^{\ast}_{\textbf{ABS}_l}$
\EndFor
\State $\mathcal{T} := \{{\bf ABS}_l \mid l \in \{1,\ldots,|\mathcal{C}|\}\}$
\If{$|\mathcal{T}| \leq 2$}
\State $\mathcal{C} = \{ C_1 \}$ where $C_1 := \mathcal{T}$
\Else
\State $\mathcal{C}=$MarkovClustering$(\mathcal{T}, \alpha, \beta)$
\EndIf
\EndWhile 
\If{$||_{T_i \in \mathcal{C}_1} L(T_i) \neq  \overline{||{}_{T_i \in \mathcal{C}_1}L_m(T_i)}$ }
\State $\Sigma_1:=\cup_{T_i \in C_1}\Sigma_{T_i}$, where $\Sigma_{T_i}$ is $T_i$'s event set
\State $L_m({\bf CO}_{|J|})=\sup C( (||{}_{T_i \in C_1}L_m(T_i)) || \Sigma^*_1 )$
\EndIf
\end{algorithmic}
\end{algorithm}

{\bf Lines~1--3:} 
Synthesize nonblocking (and maximally permissive) decentralized supervisors $\textbf{SUP}_i$ ($i \in I$) for each specification $E_i$ according to (\ref{gei}) and (\ref{dec.sup}).

{\bf Lines 4--6:}
If there are no more than two decentralized supervisors, simply treat them as one cluster (denoted by $C_1$).

{\bf Lines 7--8:}
 If there are at least three decentralized supervisors, Markov clustering is used to group the decentralized supervisors ${\bf SUP}_i$ ($i \in I$) into disjoint clusters $C_l$ ($l \geq 1$).
This is done by the function MarkovClustering, which is adapted from \cite{DSM,DMM,goorden2019structuring,Steward1981DSM}. While we refer the technical details to \cite{DSM,DMM,Steward1981DSM} due to space limit, we outline the main steps of this function with adaptations to our abstraction-based approach.
\begin{enumerate}
\item [(i)] 
 Construct a binary matrix ${\bf P}$ (whose entries are 0 or 1) called the {\em domain mapping matrix} (DMM) \cite{DMM} to record  whether a plant component and a specification have shared events\footnote{Since one decentralized supervisor enforces one specification, in DMM recording the specification information suffices for the purpose of clustering decentralized supervisors.}.
${\bf P}({\it i,k})=1$ if specification $E_i$ and plant component ${\bf G}_k$ have shared events; otherwise ${\bf P}({\it i,k})=0$.

\item [(ii)] 
Multiply ${\bf P}$ with its transpose ${\bf P^\top}$ to get a square matrix, 
called the {\it dependency structure matrix}  (DSM) \cite{DSM,Steward1981DSM}: ${\bf P_{D}=P\times P^\top}.$
${\bf P_D}(i,h)=J$ means that there exist  $J$ plant components that have shared events with both specifications $E_i$ and $E_h$ ($i,h \in I$).
\item [(iii)] 
Convert ${\bf P_D}$ into a (column) stochastic matrix $\textbf{M}$ by normalizing the columns of ${\bf P_D}$: ${\bf M}(i,h) = \frac{  {\bf P_{D}}(i,h) } { \sum\limits_{i} {\bf P_{D}}(i,h)  }.$
Thus each entry ${\bf M}(i,h)$ represents the transition probability from $i$ to $h$.
\item [(iv)]
Perform the {\it expansion} operation to the stochastic matrix {\bf M} by the following iterative scheme with parameter $\alpha$: ${\bf M}_{p+1} =  ({\bf M}_{p})^{\alpha}$, with ${\bf M}_{0} = {\bf M},$  where the expansion is the powers of matrix {\bf M} meaning the random walker does $\alpha$ jumps in each iteration {\it k}.  The choice of $\alpha$ usually needs to be adjusted based on specific datasets and experiments. Generally speaking, the choice of $\alpha$ can consider dataset size, dataset density, and clustering objective.
Through experiments and adjustments, the most suitable value $\alpha$ for the current dataset can be found to achieve ideal clustering results.
According to \cite{Wilschut2017}, 
the value of 
$\alpha$ is typically set to be $2$ for the sake of algorithm stability.
When $\alpha = 1$, the clustering process may diverge, while when  $\alpha=3$, the clustering has less influence on the
 results (for more details refer to \cite{Wilschut2017}). 
For this reason, $\alpha$ is not a tunable parameter and we simply set $\alpha =2$.
\item [(v)]
Perform the {\it inflation} operation to strengthen the probability $\textbf{M}_{p+1}(i,h)$ in ${\bf M}_{p+1}$ with parameter $\beta$:
\begin{eqnarray}\label{inflation}
{\bf M}_{p+1}(i,h) := \frac{ ({\bf  M}_{p+1}(i,h))^\beta } { \sum\limits_{\it i=1}^{\it |I|} ({\bf  M}_{p+1}(i,h))^\beta  }.
\end{eqnarray}
With this inflation operation, the high-value transition probabilities are increased while low-value ones are decreased (i.e. effect of polarization). The rate of increasing/decreasing can be controlled by the parameter $\beta$, which is tunable. The bigger $\beta$ results in more clusters with smaller cluster sizes. Thus by adjusting this parameter $\beta$, we can obtain different clusters with more or fewer decentralized supervisors. 
\end{enumerate}

Denote by $\tilde{{\bf M}}$ the resulting matrix after step~(v). Decentralized supervisors ${\bf SUP}_i$ and ${\bf SUP}_h$ are grouped into the same cluster if the $(i,h)$-entry of $\tilde{{\bf M}}$ is nonzero, i.e. $\tilde{{\bf M}}(i,h) \neq 0$. 
Thus in this way, the MarkovClustering function groups the decentralized supervisors ${\bf SUP}_i$ ($i \in I$) into a set $\mathcal{C}$ of clusters based on $\tilde{{\bf M}}$. 

{\bf Lines 9--16:}  
If there are more than two clusters, verify for each cluster $C_l$ ($l \in [1,|\mathcal{C}|]$) if the belonging decentralized supervisors are nonconflicting ($T_i = {\bf SUP}_i$ on line~11):
\begin{align} ||_{\textbf{SUP}_i \in C_l} L(\textbf{SUP}_i) =  \overline{||{}_{\textbf{SUP}_i \in C_l}L_m(\textbf{SUP}_i)}. \end{align}
Namely, check if the synchronous product of the belonging decentralized supervisors  is nonblocking.
If so, ${\bf CLU}_l$ on line~12 is nonblocking. Otherwise, design a coordinator $\textbf{CO}_l$ to resolve the conflict by line~15, 
where $\Sigma_l$ on line~14 is the union of the event sets on which the decentralized supervisors in $\mathcal{C}_l$ are defined. According to \cite{Feng2008}, this coordinator $\textbf{CO}_l$ renders cluster $\mathcal{C}_l$ nonconflicting: $(||_{\textbf{SUP}_i \in C_l} L(\textbf{SUP}_i)) || {\bf CO}_l =  \overline{(||{}_{\textbf{SUP}_i \in C_l}L_m(\textbf{SUP}_i)) || {\bf CO}_l}. $
Therefore, ${\bf CLU}_l$ on line~16 is nonblocking.

{\bf Line 17:} 
Abstract the nonblocking cluster ${\bf CLU}_l$ into an abstracted cluster ${\bf ABS}_l$. Specifically, let $\Sigma_{{\bf CLU}_l}$ and $\Sigma_{{\bf ABS}_l}$ be respectively the event sets of ${\bf CLU}_l$ and ${\bf ABS}_l$. Then abstracted cluster/abstraction ${\bf ABS}_l$ is such that $L_m(\textbf{ABS}_l)=\textit{P}_l \left( L_m(\textbf{CLU}_l)\right)$, where $\textit{P}_l : \Sigma^{\ast}_{\textbf{CLU}_l} \rightarrow \Sigma^{\ast}_{\textbf{ABS}_l} $ is a natural observer. Algorithms for determining if $P_l$ is a natural observer, and if not, converting $P_l$ into a natural observer by finding proper $\Sigma_{{\bf ABS}_l}$ can be found in \cite{Feng2010}.

{\bf Lines 18--22:}
If there is no more than two abstracted clusters ${\bf ABS}_l$, simply treat them as one cluster  (denoted by $C_1$ on line~20). Otherwise (i.e. at least three abstracted clusters), Markov clustering is used again to group these ${\bf ABS}_l$ in the same way as described in steps~(i)-(v) above (here instead of decentralized supervisors, abstracted clusters are grouped). Then lines~9--22 is repeated until there is a single cluster $C_1$ of abstracted clusters.

{\bf Lines 23--25:}
 When there is a single cluster $C_1$, check directly if the automata in this cluster are nonconflicting. If not, design the final coordinator ${\bf CO}_{|J|}$ to resolve this top-level conflict (lines~24, 25).

By {\bf Algorithm~1} above, we obtain a set of decentralized supervisors 
$\textbf{SUP}_i$ ($i \in I$) and a set of coordinators ${\bf CO}_j$ ($j \in J$) (the number of coordinators is case-dependent).  The structure of {\bf Algorithm~1} is similar to that in \cite{Feng2008}, with the key novelty of applying Markov clustering to automatically group decentralized supervisors (line~8) or higher-level abstractions (line~22). The complexity of {\bf Algorithm~1} is also the same as that in \cite{Feng2008}.  Our theoretical result is the following.

\begin{theorem}\label{thm1}
The outputs of {\bf Algorithm~1} --- $\textbf{SUP}_i$ ($i \in I$) and ${\bf CO}_j$ ($j \in J$) ---  collectively solve the Nonblocking Hierarchical Control Problem; namely (\ref{co}) holds.

\end{theorem}

Theorem~\ref{thm1} asserts that the resulting hierarchy of decentralized supervisors and coordinators collectively achieves global nonblocking controlled behavior. This result is the same as that in \cite{Feng2008}, and the key to ensure nonblockingness is the natural observers at line~17 for abstraction.\footnote{If in addition to natural observer, property of {\em output control consistency} \cite{Feng2008} or {\em local control consistency} \cite{Schmidt2011} is imposed on projections, the resulting hierarchical conrolled behavior is not only globally nonblcoking but also maximally permissive.} On the other hand, our introducing Markove clustering for automatic grouping is shown to preserve this correctness result (as is expected). For this reason and also the space limit, the proof of Theorem~\ref{thm1} is omitted, but may be referred to \cite{longversion}.

\begin{remark}
We discuss a general guideline for tuning the parameter $\beta$. Note that $\beta$ should neither be too small nor too large. A very small $\beta$ generally results in few big clusters with many components; consequently the nonconflicting checking and coordinator design in each such big cluster become computationally inefficient. On the other hand, a very large $\beta$ can fail to cluster in the sense that each component is a cluster of its own (i.e. singleton partition of the set of components); consequently the condition on Line~9 may never be met and {\bf Algorithm~1} would fail to terminate. In view of the above, one may start with an arbitrary value of $\beta$ and implement {\bf Algorithm~1} such that $|\mathcal{C}|$ on Line~9 is printed out and the time consumed on Line~11 is recorded. If Line~11 consumes exceedingly long time, this means that $\beta$ may be too small and should be increased. On the other hand, if the same value of  $|\mathcal{C}|$ is printed out consecutively, this means that $\beta$ is too large and should be decreased. Tune $\beta$ in such a way until Line~11 is efficiently executed while the value $|\mathcal{C}|$ on Line~9 decreases monotonically.
\end{remark}

\section{Case study}
\label{chap:4}
\begin{figure}[t]
	\centering

	\includegraphics[width=90mm]{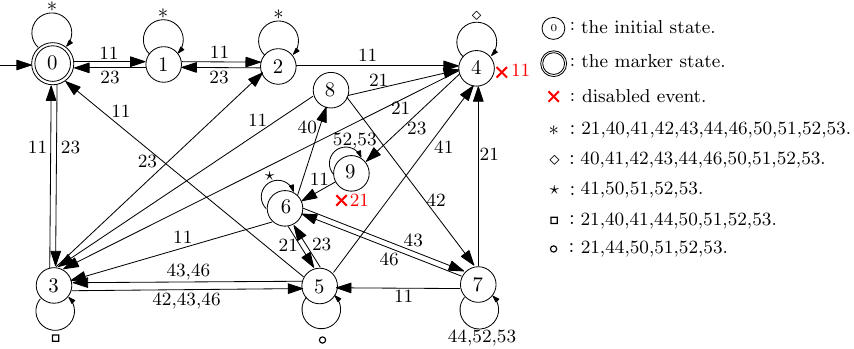}
	\caption{10-state coordinator {\bf CO} ($\beta = 4$)}
 	\label{fig:AGVco10}
\end{figure}
In this section, we demonstrate the effectiveness of our proposed Markov clustering based approach on one benchmark case study: 
automatic guided vehicles (AGVs). 
The AGV system consists of five automatic guided vehicles {\bf AGV1}$,...,${\bf AGV5} serving a manufacturing workcell.  This workcell consists of two input stations {\rm IPS1, IPS2}; three workstations {\rm WS1, WS2, WS3}; one completed parts station {\rm CPS}; and four shared zones Z1, Z2, Z3, Z4 for the AGVs. Detailed description of the system and the automata of the AGVs are referred to \cite{Feng2008}. 
For this system, the following control specifications are imposed:
\begin{itemize}
\item Each of the four shared zones Z1, Z2, Z3, Z4 should be occupied by at most one AGV at a time.
\item Only one of {\bf AGV1, AGV2} can be loaded at a time in two input stations {\rm IPS1, IPS2}.
\item Only one part can be processed at a time by each of {\rm WS2}, {\rm WS3}, while {\rm WS1} can assemble just two parts (a Type1 and a Type2) at a time into a complete part. Three workstations must be protected against overflow and underflow. 
\end{itemize}
The above requirements can be modeled by 9 specification automata (\cite{Feng2008}): $\textbf{Z}_1, \textbf{Z}_2,\textbf{Z}_3, \textbf{Z}_4, \textbf{WS}_{13}, \textbf{WS}_{14}, \textbf{WS}_{2}, \textbf{WS}_{3}, \textbf{IPS}.$
In the following, we apply our proposed  procedure. Due to space limit, details are referred to \cite{longversion}. Specific to Markov clustering, first by line~8 in {\bf Algorithm~1} we need to cluster the nine decentralized supervisors (computed by line~3 in {\bf Algorithm~1} for the nice specifications). To this  end, the following stochastic matrix is constructed: 

\vspace{-0.7cm}
\[
\textbf{M} = \left(
\begin{smallmatrix}
	0.2500  &  0.1250  &  0.1111     &    0          &    0          &    0            & 0.1667     &  0.1111     &  0.2500 \\
    0.1250  &  0.2500  &  0.1111     &      0       & 0.1667     &       0         &  0.1667    &  0.1111    &   0.1250 \\
    0.1250  &  0.1250  &  0.2222    &  0.1429   &     0          &   0.1429   &      0         & 0.2222      & 0.1250 \\
         0     &    0        & 0.1111        & 0.2857  &  0.1667    &  0.2857     &    0          & 0.1111      &      0 \\
         0   & 0.1250     &    0          &  0.1429    & 0.3333     &  0.1429     &  0.1667   &       0        &   0 \\ 
         0     &    0         & 0.1111      & 0.2857   &  0.1667     & 0.2857    &     0           &  0.1111   &      0 \\
    0.1250 &   0.1250    &     0        &     0        &   0.1667    &     0     &  0.3333         &     0         &  0.1250  \\
    0.1250 &   0.1250  &  0.2222    & 0.1429    &     0         &  0.1429     &    0           &  0.2222    &  0.1250  \\ 
    0.2500  &  0.1250  &  0.1111     &    0        &   0            &   0            & 0.1667     &  0.1111    &  0.2500 \\
\end{smallmatrix}
\right).
\]
\vspace{-0.7cm}
\renewcommand{\arraystretch}{1}


This {\bf M} is used as the input of the Markov clustering algorithm. While we fix parameter $\alpha=2$, the parameter $\beta$ is varied for different values ($2.5 \sim 10$) as shown in Table~I. Observe that larger $\beta$ results in more clusters (with fewer decentralized supervisors in each cluster). The third column in Table~I shows that for all tried values of $\beta$, there is only one higher-level coordinator needs to be computed to ensure global nonblocking behavior; however, small values of $\beta$ cause the coordinator to have larger state sizes. 
 On the other hand, larger values of $\beta$ often require more execution times of the algorithm (see the fourth column of Table~I),\footnote{Time is measured on a personal computer with i7 CPU and 8G memory.} as more layers of computations are needed.


\renewcommand{\arraystretch}{0.5}
\begin{table}[H]
\resizebox{1\columnwidth}{!}{
    \centering
\label{tab:AGV_beta}
    \begin{tabular}{|c|c|c|c|c|}
    \hline
        
        \begin{tabular}{c}
          $\beta$
        \end{tabular}
        &  
        \begin{tabular}{c}
          Numbers of clusters for\\
           decentralized supervisors
        \end{tabular}
        &
        \begin{tabular}{c}
           
          Coordinator\\
        (state size)
        \end{tabular}
        &
        \begin{tabular}{c}
           
           Execution\\
         time [s]
        \end{tabular}\\
        \hline
        \hline
        2.5  & 2&  $\textbf{CO}$ (64) &  13.164 \\
        \hline
		3    & 3 & $\textbf{CO}$ (51) &  17.938 \\
         \hline
		3.5  &  3 & $\textbf{CO}$ (27) &  17.876 \\
        \hline
		4    &  4 & $\textbf{CO}$ (10) &  19.120 \\
  \hline
		4.5  & 4 & $\textbf{CO}$ (10) &  24.806 \\
  \hline
		5    & 4 & $\textbf{CO}$ (10) &  30.135 \\
  \hline
		5.5  &5 &  $\textbf{CO}$  (10) &  22.143 \\
  \hline
		6    & 5 & $\textbf{CO}$ (10) &  22.070\\
  \hline
		6.5  & 5  &  $\textbf{CO}$ (10) &  22.225 \\
  \hline
		7    & 5  &  $\textbf{CO}$ (10) &  22.581 \\
  \hline
    	7.5  &  5 &  $\textbf{CO}$ (10) &  22.220 \\
     \hline
		8    &6  & $\textbf{CO}$ (10) &  90.650  \\
  \hline
		8.5  & 6  & $\textbf{CO}$ (10) &  98.517  \\
  \hline
		9    & 6 & $\textbf{CO}$ (10) &  106.299  \\
  \hline
		9.5  & 6 & $\textbf{CO}$ (10) &  115.805   \\
  \hline
		10   &  6 & $\textbf{CO}$ (10) &  123.901  \\

        \hline
    \end{tabular}}
     \caption{Tuning parameter $\beta$ for different clusters}
\end{table}
\vspace{-0.6cm}

As an example, consider $\beta = 4$. The resulting $4$ clusters of decentralized supervisors are shown in Table~II. Also the 10-state higher-level coordinator is displayed in Fig.~\ref{fig:AGVco10}.\footnote{In \cite{Feng2008} by detailed analysis of the structure of the AGV system, a higher-level coordinator of 7 states is found. Our automatically resulted coordinator is fairly close to the handcrafted one in \cite{Feng2008}.} 
\renewcommand{\arraystretch}{1.5}
\begin{table}[h]
	\centering
 \caption{Clusters of decentralized supervisors ($\beta = 4$)}
\label{tab:AGV_C}
	\begin{tabular}{|c|c|c|c|}
		\hline
		Cluster 1 &Cluster 2&Cluster 3 &Cluster 4\\
		\hline
		\hline
		$\textbf{SUP}_{\textbf{WS}_{13}}$ & $\textbf{SUP}_{\textbf{Z}_4}$ & $\textbf{SUP}_{\textbf{Z}_3}$ & $\textbf{SUP}_{\textbf{Z}_1}$ \\
		                                                          & $\textbf{SUP}_{\textbf{WS}_{14}}$ & $\textbf{SUP}_{\textbf{WS}_{3}}$ & $\textbf{SUP}_{\textbf{Z}_2}$\\
		                                                            &                                                &                                                          & $\textbf{SUP}_{\textbf{WS}_{2}}$ \\
		                                                           &                                                &                                                           &  $\textbf{SUP}_{\textbf{IPS}}$\\
		\hline
	\end{tabular}
\end{table}
\renewcommand{\arraystretch}{1}

Finally it is confirmed that the joint behavior of the 10-state coordinator and the nine decentralized supervisors is not only nonblocking, but also in this case maximally permissive (thus equivalent to the controlled behavior of the monolithic supervisor). Therefore for this case study, our fully automated approach successfully solve the nonblcoking hierarchical control problem and moreover the resulting coordinator is fairly simple (the derivation of which does not need knowing or analysis of system's structure).

\section{Conclusions}
 We have proposed a fully automated, effective, and flexible hierarchical synthesis procedure with Markov clustering for large-scale DES, with no need of knowing or analyzing system structures.
We have proved that the resulting hierarchy of supervisors and coordinators collectively achieves global nonblocking (and maximally permissive) controlled behavior under the same conditions as those in the existing abstraction-based approach. Our approach  additionally provides flexibility in tuning the cluster sizes by adjusting a single parameter used in Markov clustering. Our approach is modular which can result in smaller and thus more comprehensible decentralized supervisors/coordinators Finally, a benchmark case study has been conducted to demonstrate the effectiveness of our approach.


\bibliographystyle{plain}        
\bibliography{autosam-2/autosam2}           



\end{document}